\documentclass[9pt,twocolumn,twoside]{pnas-new}
\templatetype{pnasresearcharticle} % Choose template 
\title{Policing Chronic and Temporary Hot Spots of Violent Crime: A Controlled Field Experiment}

\author[a,1]{Dylan J. Fitzpatrick}
\author[a]{Wilpen Gorr} 
\author[b]{Daniel B. Neill}

\affil[a]{Carnegie Mellon University}
\affil[b]{New York University}

\leadauthor{Fitzpatrick} 

\correspondingauthor{\textsuperscript{1}To whom correspondence should be addressed. E-mail: djfitzpa@alumni.cmu.edu}


\begin{abstract}
Hot-spot-based policing programs aim to deter crime through increased proactive patrols at high-crime locations. While most hot spot programs target easily identified chronic hot spots, we introduce models for predicting temporary hot spots to address effectiveness and equity objectives for crime prevention, and present findings from a crossover experiment evaluating application of hot spot predictions to prevent serious violent crime in Pittsburgh, PA. Over a 12-month experimental period, the Pittsburgh Bureau of Police assigned uniformed patrol officers to weekly predicted chronic and temporary hot spots of serious violent crimes comprising 0.5 percent of the city’s area. We find statistically and practically significant reductions in serious violent crime counts within treatment hot spots as compared to control hot spots, with an overall reduction of 25.3 percent in the FBI-classified Part 1 Violent (P1V) crimes of homicide, rape, robbery, and aggravated assault, and a 39.7 percent reduction of African-American and other non-white victims of P1V crimes. We find that temporary hot spots increase spatial dispersion of patrols and have a greater percentage reduction in P1V crimes than chronic hot spots but fewer total number of crimes prevented. Only foot patrols, not car patrols, had statistically significant crime reductions in hot spots. We find no evidence of crime displacement; instead, we find weakly statistically significant spillover of crime prevention benefits to adjacent areas. In addition, we find no evidence that the community-oriented hot spot patrols produced over-policing arrests of minority or other populations.
\end{abstract}

\begin{document}

\maketitle
\thispagestyle{firststyle}
\ifthenelse{\boolean{shortarticle}}{\ifthenelse{\boolean{singlecolumn}}{\abscontentformatted}{\abscontent}}{}

% If your first paragraph (i.e. with the \dropcap) contains a list environment (quote, quotation, theorem, definition, enumerate, itemize...), the line after the list may have some extra indentation. If this is the case, add \parshape=0 to the end of the list environment.
\dropcap{I}n recent years, police agencies have demonstrated an increased willingness to devote resources to proactive policing strategies for preventing crime \citep{Weisburd2004}. In contrast to reactive policing, which prioritizes responding to 911 emergency requests for police response and investigating crimes that have already been committed, proactive policing aims to identify predictable high-crime patterns and to take actions in regard to those patterns to prevent crimes \citep{NRC2004}. Most recently, in the U.S., discussion of the potential benefits and harms of proactive policing has become part of a wider debate about the role of police and their impacts on the communities they serve, sparked by multiple instances of police brutality against minorities and widespread protests against systemic racism in policing as part of the Black Lives Matter movement \citep{Thurston2018}.

Here we present results from a controlled experiment of proactive policing which contributes several pieces of evidence to this debate, quantifying the crime prevention benefits of proactive patrols targeting predicted “hot spots” of serious violent crime, and confirming that certain undesirable outcomes (over-policing arrests of racial minorities) did not occur in the study period. We emphasize that these impacts are critically dependent on the community-oriented approach taken by police patrolling predicted hot spots, while aggressive policing practices may substantially harm individuals and communities. In the discussion below, we note other critical issues to be addressed while evaluating the potential impacts of proactive policing, as well as describing a set of best practices intended to maximize benefits and minimize harms.

\subsection*{Crime prevention at crime hot spots}
While certain high-crime patterns may involve specific persons, our research concerns high-crime areas rather than individuals. A primary police intervention for preventing crimes in such areas is frequent, targeted patrols with police presence deterring criminals from committing crimes \citep{Braga2019, Sherman1995, Mohler2015, Mohler2011}. The Part 1 Violent (P1V) crimes of homicide, rape, aggravated assault, and robbery are the most serious offenses occurring in a city and are the target crimes for our study (excluding domestic incidents).

An established finding in the criminology literature is that crimes do not occur uniformly across cities, but instead concentrate in micro-places, or ``hot spots.'' Hot spots range from single intersections or storefronts to a few city blocks and represent optimal locations for committing crimes \citep{Sherman1989, Block1995, Eck1995, Brantingham1999, Weisburd2015}. Crime hot spots make up a small fraction of the total area of a city yet produce disproportionately many calls for police response to a range of violent, property, and disorder crimes. 

Enforcement actions such as stops, searches, and arrests can prevent crime by directly incapacitating potential offenders \citep{Weisburd2004}. Incapacitation may have immediate effects on crime \citep{Wyant2012} or longer-term effects if prolific offenders are removed from a community. However, it has been shown that aggressive policing practices, including frequent stops, summonses, and arrests for low-level crimes, have adverse impacts on community health \citep{Geller2014}, police-community relations \citep{NAS2018}, and racial equity. Minorities can suffer from ``over-policing'' with aggressive enforcement. An example is the ``broken windows'' approach to policing that advocates zero tolerance to disorder behavior in the belief that it would reduce crime. Strict enforcement leads to disproportionate harassment and arrest of minority persons and erodes public trust in police \citep{Kamalu2018}.

Guardian actions, such as increased police presence through patrols and community-policing methods, deter crime by removing opportunities to commit crime and increasing the perceived risk of crime commission, even though patrols generally do not involve direct contact with potential offenders \citep{Sherman1995, Loughran2011}. A substantial body of evidence reveals that sending proactive patrols to selected crime hot spots for guardianship reduces crime in those areas, but the amount of observed crime reduction varies considerably across studies to date \citep{NAS2018}. A meta study on guardian actions in the form of community-oriented policing and problem solving versus enforcement actions in hot spots found guardianship to have larger effect sizes \citep{Braga2015}. Spatial diffusion of patrol benefits is supported by several studies showing that the deterrent effect from police presence diffuses into areas near patrolled hot spots \citep{Clarke1994,Weisburd2006, Telep2014}.

During the 1980s and 1990s, foot patrols were used in large areas of cities, such as entire police beats. Such patrols improved the public’s perceptions of safety, livability, and positive attitudes toward police but showed no crime reduction effects \citep{Bowers2004, Cordner1994}. Seminal work by Ratcliffe et al. in Philadelphia \cite{Ratcliffe2011}, followed by replications by Novak et al. in Kansas City, MO \cite{Novak2016} and Piza in Newark, NJ \cite{Piza2018}, found significant reductions in serious violent crimes with controlled experiments applying saturation (full-time) foot patrols to micro-places. Saturation patrols are costly, so Haberman and Stiver \cite{Haberman2019} applied only two hours of foot patrol per shift in Dayton, OH, and did not find a significant reduction in violent crime.  In contrast, our paper provides evidence that foot patrols achieve statistically significant reductions in serious violent crime, using only the discretionary time of regular patrol officers for short-duration patrols.

\subsection*{Chronic and temporary hot spots}
Most hot spot studies focus on the impact of patrolling a fixed set of static hot spots pre-selected using simple tabulations of crimes by location, crime density maps, crime intelligence, or expert knowledge from police. These chronic hot spots most often are large commercial areas and adjacent residential areas, and result in a high concentration and volume of crime that persist over the course of years and even decades \citep{Gorr2015, Weisburd2011}. However, a growing body of evidence finds that other hot spots are dynamic, with crime clusters exhibiting on-and-off behavior over short periods of time \citep{Herrmann2015, Mohler2015, Gorr2015, Gorr2017}. Temporary hot spots represent flare-ups of crime resulting from short-term phenomena, such as escalating gang disputes or repeat crimes by transient individuals. Temporary hot spots tend to be widely distributed throughout a city including small commercial areas and residential areas \citep{Gorr2015}. Our research uses prediction and selection of chronic and temporary hot spots as two policy levers for police to address effectiveness and equity of crime prevention.

\subsection*{Effectiveness and equity objectives}

Police command staff must weigh resource and manpower constraints when making decisions about how to distribute limited patrols throughout a city. While most research on crime hot spots has used effectiveness in forecasting and reducing crimes as the sole objective, we also include equity in the distribution of crime prevention services by police across a city as an additional objective. Equity is an important objective for public goods, and especially for primary goods such as public safety \citep{Culyer2001}.

For evaluating effectiveness of hot spot forecasting methods, we use plots of percentage of a city’s crimes predicted in hot spots versus percentage area of a city in hot spots \citep{Johnson2009, Mohler2014, Gorr2015, Chainey2008}, here termed ``prediction tradeoff curves''. These curves are upward sloping with marginally decreasing returns from increased patrols, and are directly useful for comparing accuracy of alternative prediction models, incorporating resource limitations of police, and showing tradeoffs between effectiveness and equity objectives. See \textit{SI: Evaluation of Prediction Methods on Historical Data} for example curves for our models.

The underlying principles of equity are that “like needs should receive like attention and resources… and greater needs should receive greater attention and resources… [where the thing needed] is instrumental, a necessary condition if a more ultimate objective is to be met” \citep{Culyer2001}. Equity in crime prevention across a city, as measured by the spatial dispersion of areas selected as hot spots across time, is an important consideration for police \citep{NAS2018, Gorr2015, Gorr2017}. The ultimate objective for crime prevention is a high level and equal distribution of public safety for all \citep{Weisburd2012}. Experiments, such as the one reported in this paper, confirm the instrumentality of proactive patrols in hot spots for increasing public safety in all parts of the city where target crimes are predictable. See \textit{SI: Equity in the Allocation of Crime Prevention Services by Police} for further explanation of these principles applied to crime prevention and how we developed corresponding tools for policy makers to improve equity.

As a basis for measures of equity, we use a hot spot program’s ``footprint,'' its spatial distribution of hot spots over a year that have been patrolled at least for a week. With an equal number of chronic and temporary hot spots in a program, we expect the footprint of temporary hot spots to be much larger and widely distributed than that for chronic hot spots. We also use an entropy index to summarize spatial dispersion of chronic versus temporary hot spots.

\subsection*{Crime forecasting models and methods}
Although simple methods can accurately identify chronic hot spots, the emerging crime patterns of temporary hot spots require sophisticated spatio-temporal forecasting models. While such models are intended to predict temporary hot spots, they also predict chronic hot spots if using crime volume as the dependent variable. Instead, our temporary hot spot model forecasts deviations from baseline crime levels, thereby yielding temporary hot spots and occasionally flare-ups at chronic hot spots. 

To date, predictive policing studies have focused on a single method for forecasting hot spots and have not included a rigorous model selection step in which multiple forecasting methods for hot spots are compared. Our study uses a grid of Pittsburgh consisting of cells 500 feet on a side as potential hot spots. Through an analysis of five years of historical reported crime incidents and 911 data ending in 2016, we selected forecast models that most accurately forecasted chronic and temporary hot spots of P1V crimes from a set of ten candidate models from the crime hot spot and forecasting literature (See \textit{Materials and Methods}). Most accurate for chronic hot spots was a simple 12-month moving average of serious violent crimes by grid cell. Most accurate for temporary hot spots was a single-layer, fully-connected neural network model with time-lagged dependent and independent variables made up of selected crimes and 911 call for service counts from a two-year moving window of data. 

\subsection*{Field Study}
Pittsburgh, PA is a city of 300,268 population with 64.9\% white (non-Hispanic), 22.8\% African American, 5.7\% Asian, and 2.3\% Hispanic racial composition. The Pittsburgh Bureau of Police (PBP) has approximately 900 sworn officers distributed across six police zones, each with a police station, commander, and officers. Prior to our field study, PBP maintained a policy that uniformed patrol officers conduct proactive patrols informed by experience and recent crime events to prevent crimes and protect citizens. For such patrols, officers used discretionary time when not responding to 911 calls for service or other official duties. PBP policy for proactive patrols required community-oriented policing methods such as engaging with local citizens, avoiding enhanced enforcement actions such as zero-tolerance policing \citep{PBP2009}. Officers were free to choose between car and foot patrol. Foot patrol used ``park and walk'' so that parked marked police cars provided additional police presence.

PBP set the goal that each officer assigned a hot spot spend an average of 15 minutes per shift on proactive patrol at the hot spot, with officers assigned to hot spots across all three shifts and seven days per week (45 minutes of patrol per hot spot per day). The result was a small-scale program with 36 hot spots for the city consisting of six hot spots (three chronic and three temporary) for each of Pittsburgh’s six police zones with 0.5 percent of the area of the city under treatment at any given time. During the study year, actual treatment duration averaged 35.5 minutes per day for both chronic and temporary hot spots, less than the targeted 45 minutes.

We evaluated the hot spot program through a controlled field study motivated by crossover trials from the field of public health. Using historical data on serious violent crime frequency, we divided each of the six police zones into two halves of roughly equal area and counts of historical violent crimes to minimize spillover contamination of control hot spots by proactive patrols at treatment hot spots. Residual crime-deterrence effects after patrols leave an area are typically only one or two hours \citep{Koper1995, Telep2014}, thus temporal spillover effects from treatment in previous weeks to control hot spots were expected to be negligible. The areas exposed to treatment were initially randomized and then alternated on a weekly basis between the halves of each police zone, so that no grid cell was selected for treatment two weeks in a row. Predicted hot spots in the ``control'' partitions for a given week were not provided to PBP; thus, control hot spots had policing as usual, including reactive policing in response to 911 emergency calls for service and a limited number of police-designed proactive patrols that existed before and during the experiment.

February through October 2017 included first a pilot implementation in one of the six zones, followed by sequential rollout of the hot spot program in the remaining five zones, with a burn-in period to increase patrol numbers to impactful levels. The field study ran for 12 months from November 6, 2017 through November 4, 2018. During the study year, there were a total of 37 homicides, 56 rapes, 686 robberies, 682 aggravated assaults, 1230 burglaries, 6394 larcenies, and 772 motor vehicle thefts (exclusive of domestic violence and retail crimes).

\section*{Results}
\label{sec:results}

\subsection*{Per-cell reductions in crime volume from proactive patrols}
Policing-as-usual for patrol officers in Pittsburgh, while mainly consisting of responding to 911 calls for emergency services, also included a number of police-generated proactive patrols. We therefore expected that control hot spots, known to the researchers but not provided to PBP, would have some police-generated proactive patrols. Overall, control hot spots had proactive patrols in numbers equal to 14.8 percent of patrols conducted in treatment hot spots (1.9 vs. 12.9 patrols per hot spot per week), with 18.8 percent for chronic hot spots (2.6 vs. 13.8 patrols per hot spot per week) and 10.1 percent (1.2 vs. 12.1 patrols per hot spot per week) for temporary hot spots. Thus, the observed differences between P1V crimes in treatment and control hot spots represent the impact of targeting predicted (chronic or temporary) hot spots on a given week of the experiment, resulting in an additional 11 proactive patrols to targeted cells on average.

We employ a fixed-effects regression model to evaluate statistical significance of per-cell P1V crime reductions from the hot spot program in Pittsburgh. The units of analysis are cell-weeks, and outcomes $Y$ are Part 1 Violent (P1V) crime counts per cell. We defined a regression model (Equation \ref{eq:reg_model}) with a binary variable $T_{p,w}$ indicating whether partition $p$ was selected for treatment in week $w$, $HS_{c,w}$ indicating whether cell $c$ was selected as a hot spot in week $w$, and $D_{c,w}$ indicating the number of proactive patrols in cell $c$ during week $w$. $HS_{c,w}\times D_{c,w} \times T_{p,w}$ is then an interaction term between hot spot selection and treatment dose. Fixed effect terms for partitions and weeks are $\alpha_p$ and $\delta_{w}$ respectively, and $\epsilon_{c,p,w}$ terms are cell-specific errors. Model coefficients are estimated through ordinary least squares regression.

\begin{align}
\label{eq:reg_model}
\begin{split}
Y_{c,p,w} &= \beta_0+ \beta_1[HS_{c,w}\times D_{c,w}\times T_{p,w}] 
\\&+ \beta_2 HS_{c,w} + \alpha_p + \delta_{w} + \epsilon_{c,p,w}
\end{split}
\end{align}

To minimize spillover of deterrence effects across treatment and control boundaries, cells in the control partition but directly adjacent to treatment hot spots were dropped from the analysis. Similarly, cells in the treatment partition that were directly adjacent to control hot spots were dropped. The primary coefficient of interest is $\beta_1$, which represents the per-cell reduction in target crimes within treatment hot spots associated with each additional proactive patrol. We report results for four model variants: Model 1 examines the dose-dependent treatment effect of patrols to hot spot cells and includes week and zone-partition fixed effects. Model 2 also examines possible changes in crime volume in cells adjacent to hot spots and in non-adjacent cells. Model 3 estimates separate dose-dependent treatment effects for chronic and temporary hot spots. Finally, Model 4 estimates separate dose-dependent treatment effects for car patrols and foot patrols at chronic and temporary hot spots. We expected car patrols to be 1/3 the duration of foot patrols (5 minutes vs. 15 minutes per patrol), and therefore we compute combined patrol dose as (number of foot patrols) + (number of car patrols/3) for Models 1-3. 

A concern in estimating dose-dependent treatment effects is that dose was not randomized across hot spots. It is possible that officers may direct proactive patrols either toward or away from cells with high incidence of P1V, thereby affecting the dose-dependent treatment effect coefficient. We refute this possibility by computing the correlation between average number of proactive patrols when treated and average P1V when control for the 464 cells that are selected as both treatment and control at least once in the study period, and find null correlations for all patrol types (foot patrols vs. P1V: r=0.017; car patrols vs. P1V: r=-0.009; foot + car patrols vs. P1V: r=-0.000). See SI Table S6 for the full table of dose-dependent regression results with P1V outcomes. We report non-dose-dependent results in SI Table S7, and also provide results in SI Table S8 on Part 1 Property crimes, to demonstrate that crime prevention was not limited to the targeted P1V crimes.

\subsubsection*{Model 1} The coefficient on Hotspot × Treatment Dose is negative and statistically significant ($\beta$=-0.0009, 95\% CI=[-0.001, -0.000], \textit{p}<0.001), while the coefficient on the Hotspot indicator variable is positive and statistically significant ($\beta$= 0.0431, 95\% CI=[0.040, 0.046], \textit{p}<0.001). Taken together, these results indicate that cells selected as a hot spot experience higher crime volume on average, but treating these areas with additional proactive patrols tends to decrease crime relative to hot spots that do not receive patrols.
\subsubsection*{Model 2} We find a weakly statistically significant reduction in P1V crimes in cells adjacent to hot spots ($\beta$=-0.0014, 95\% CI=[-0.003, 0.000], \textit{p}=0.082), providing some evidence for a small degree of spatial diffusion of crime deterrence benefits to areas surrounding predicted hot spots, and no evidence for crime displacement. Note that this result represents the effect of being adjacent to a treatment hot spot regardless of treatment dose at that hot spot. We find no statistically significant change in P1V crime in cells not adjacent to hot spots ($\beta$=-0.0001, 95\% CI=[-0.001, 0.000], \textit{p}=0.895).
\subsubsection*{Model 3} The coefficients on Hotspot × Treatment Dose for both chronic and temporary hot spots are negative (indicating a reduction in crime volume from treatment) and statistically significant (chronic hot spots: $\beta$=-0.0015, 95\% CI=[-0.002, -0.001], \textit{p}<0.001; temporary hot spots: $\beta$=-0.0008, 95\% CI=[-0.001, 0.000], \textit{p}=0.024). Differences in the estimated treatment effects in chronic and temporary hot spots provide evidence that equivalent patrol protocols to chronic versus temporary hot spots may result in differing crime deterrence benefits.
\subsubsection*{Model 4} For foot patrols, the coefficients on Hotspot × Treatment Dose for both hot spot types are negative (indicating a reduction in crime volume from treatment) and statistically significant (chronic hot spots: $\beta$=-0.0031, 95\% CI=[-0.005, -0.002], \textit{p}<0.001; temporary hot spots: $\beta$=-0.0025, 95\% CI=[-0.004, -0.001], \textit{p}=0.001). In contrast, car patrols do not appear to provide P1V crime prevention (chronic hot spots: $\beta$=0.0002, 95\% CI=[0.000, 0.001], \textit{p}=0.47); temporary hot spots: $\beta$=0.0006, 95\% CI=[0.000, 0.001], \textit{p}=0.094). Since car patrols and foot patrols are positively correlated (r=0.49), dropping car patrols from Model 4 reduces the estimated dose-dependent effect size of foot patrols on P1V crime to $\beta$=-0.0027 (\textit{p}<0.001) and $\beta$=-0.0016 (\textit{p}=0.003) for chronic and temporary hot spots respectively.

\subsection*{Overall reduction in crime volume from proactive patrols} The observed differences in P1V crimes between control and treatment hot spots are reported in Table \ref{tab:P1V_reductions}. Within all hot spots, we measured 24 fewer P1V crimes in treatment hot spots relative to control hot spots, or a 25.3\% reduction in P1V crimes per hot spot cell ($n_{treatment}$ = $n_{control}$ = 1872 cell-weeks). We see a larger percentage reduction in temporary hot spots for P1V crimes, with a reduction of 33.3\% in temporary hot spots (n=936) and 23.8\% in chronic hot spots (n=936). In cells adjacent to hot spots (queen’s case), we measured an 8.6\% decrease in P1V crimes per cell in treatment areas. While the numbers of crimes reduced are small, their impact in terms of cost avoidance to society and victims is large compared to the cost of proactive police patrols. In Table 1, we compute that overall there were \$3,411,328 in crime costs avoided, while the cost of patrols was less than \$300 thousand (see \textit{SI: Evaluation of Program Impacts}). Counts and standard deviations of P1V crimes in treatment and control cells are reported in SI Table S5. We observe a statistically significant reduction in the number of African American and other non-white victims of P1V crime, with 25 (39.7\%) fewer victims in treatment vs. control cells over the study. Prevention of P1V crimes with multiple victims accounts for a higher proportional reduction in minority victims than the 25.3\% reduction in overall P1V.

\begin{table}[h!]\centering
\begin{threeparttable}
\caption[Observed differences in Part 1 violent crime counts between control treatment areas over 12 months of proactive hot spot patrols.]{Observed differences in Part 1 violent (P1V) crime counts between control areas and treatment areas over 12 months of proactive hot spot patrols (November 6, 2017 – November 4, 2018).}
\label{tab:P1V_reductions}
\begin{tabular}{lccc}
\toprule
\multirow{2}{*}{} & \multicolumn{3}{c}{Part 1 violent crimes}\\
\cmidrule(rl){2-4}
&\% Change& \# of Crimes & \shortstack{Est. Costs\\Avoided*}\\
\midrule
Chronic Hot Spots & -23.8\% & -19 &\$2,701,891 \\%\hline
Temporary Hot Spots& -33.3\% & -5 &\$709,437 \\\hline
All Hot Spots& -25.3\% & -24 &\$3,411,328 \\%\hline
\bottomrule
\end{tabular}\begin{tablenotes}
      \small
      \item *Costs per P1V crime computed based on costs to individuals and society from individual component crimes reported in \citep{McCollister2010} and inflated to 2018 dollars. Note that the cost of the proactive police patrols was less than \$300 thousand. See SI Table S9 for costs of individual component crimes.
    \end{tablenotes}
  \end{threeparttable}
\end{table}

\begin{figure}
  \setlength{\fboxsep}{0pt}
  \setlength{\fboxrule}{1pt}
 \centering
  \fbox{\includegraphics[width=4.2cm]{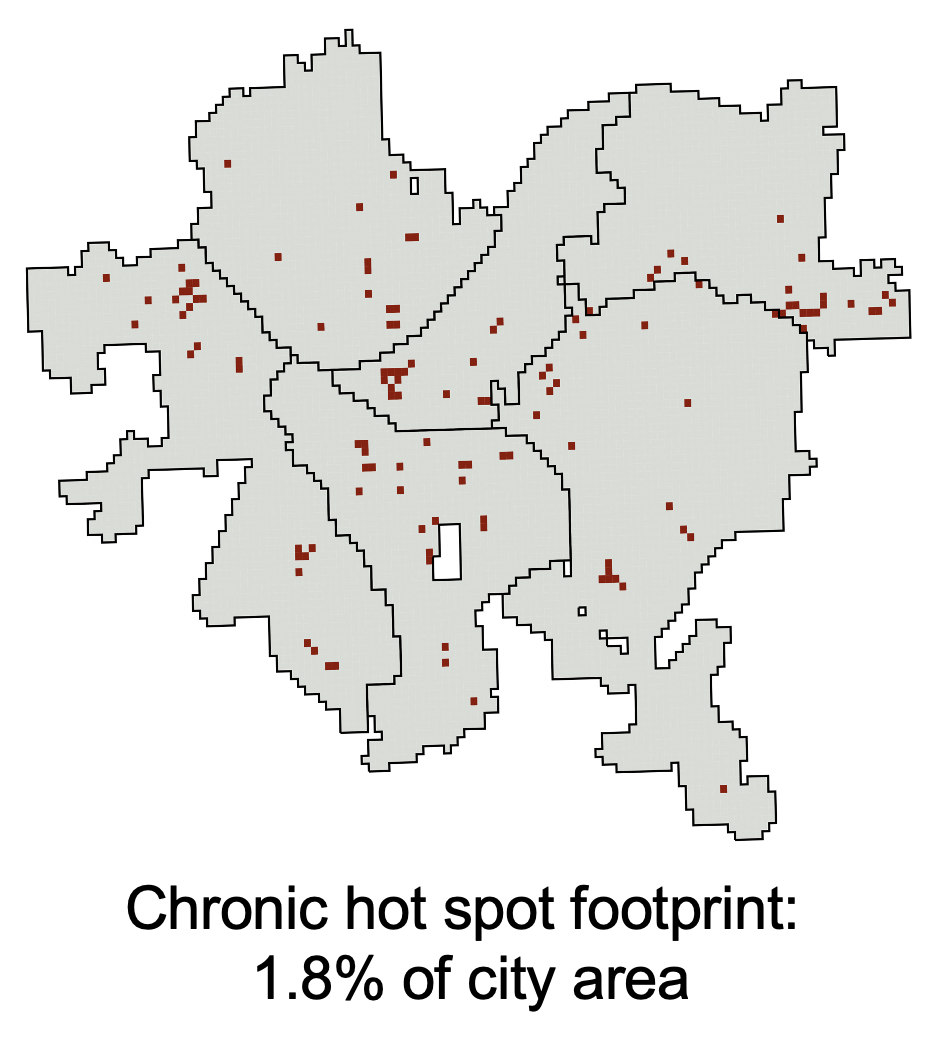}}
  \fbox{\includegraphics[width=4.05cm]{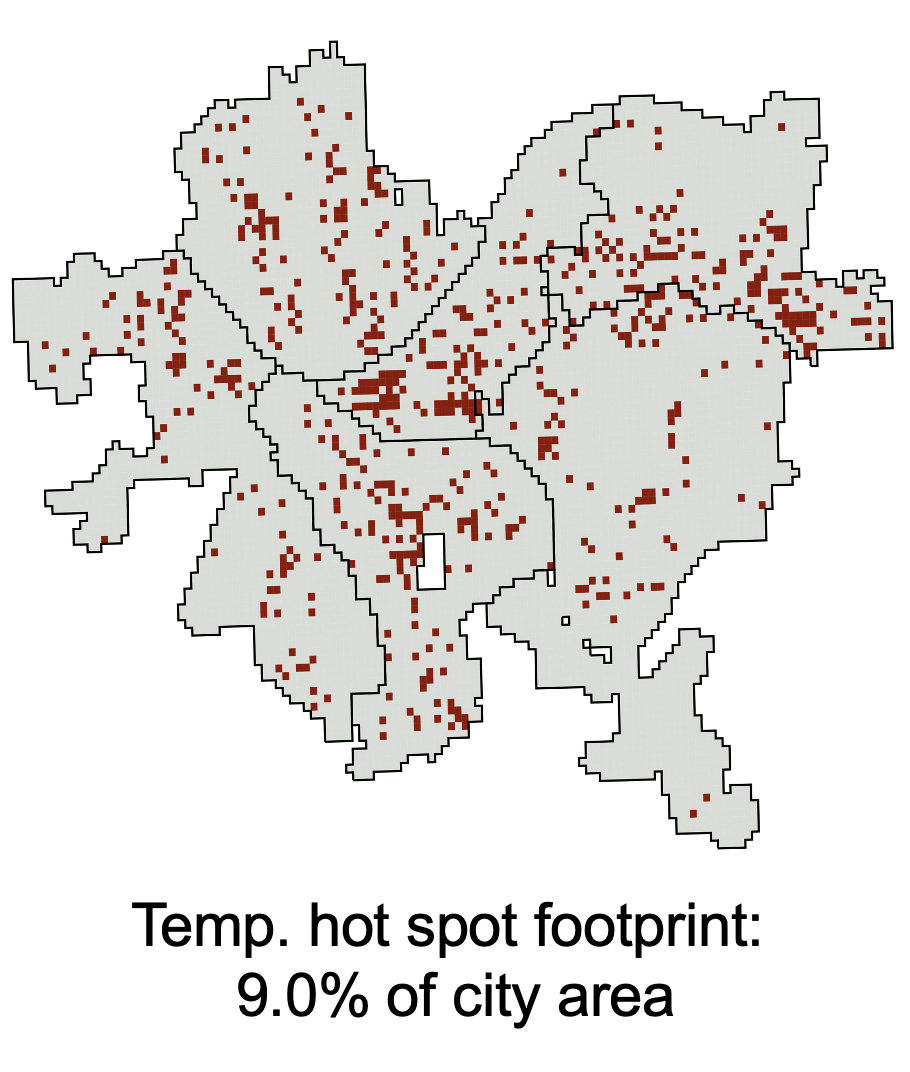}}

\caption[Hot spot footprint over 12 months for chronic and temporary hot spots.]{Hot spot footprint in Pittsburgh over 12 months for two types of hot spots, chronic hot spots predicted using a long-term moving average (left), and temporary hot spots predicted using a multilayer perceptron (right). Cells are highlighted if they were selected as a hot spot at least once in the study period. Boundaries indicate the six police zones.}
\label{fig:footprints}
\end{figure}

\subsection*{Spatial dispersion of chronic and temporary hot spots} As discussed in \textit{SI: Evaluation of Prediction Models on Historical Data}, the temporary hot spot model was selected to increase spatial dispersion of top-selected hot spots as measured by a selection entropy index. Figure \ref{fig:footprints} shows the hot spot footprints for chronic and temporary hot spots across the 12-month period of evaluation. Cells are flagged in red if they were included as a hot spot at least once in the study period. As expected, the chronic hot spots selected using a one-year moving average remained fairly static, only occurring in a small fraction of the city over 12 months (recall that there were an equal number of chronic and temporary hot spots during the study year). By contrast, temporary hot spots were spread over five times the area of chronic spots over the same period (9.0\% vs. 1.8\% of the city area).

\subsection*{Arrests during hot spot patrols} Over-policing is unnecessarily large amounts of police control and arrests, particularly for racial minority or economically disadvantaged communities \citep{Benporat2008}. Over-policing is especially a concern for “broken window” crimes that do not threaten public safety and can involve officer discretion in enforcement, such as consumption of alcohol on streets, possession of small amounts of marijuana, disorderly conduct, loitering, disturbing the peace, and jaywalking. Over-policing results from racial profiling, institutional biases ingrained in the culture or policies of a police department, or personal biases of officers. In contrast, Pittsburgh police policy for proactive patrols is to engage with the local residents and employ community-oriented policing (for example, warning of possible violent crimes and gathering information on potentially dangerous locations). To assess the impacts of the hot spot program on potential harms from over-policing, PBP crime analysts \citep{Johnson2019} tabulated the number of arrests occurring during hot spot patrols. In the time window during and 30 minutes after the over 24,000 hot spot patrols in treatment hot spots conducted during the study year, only four arrests occurred while on patrol. Two of these were on-view arrests for minor drug offenses, and the remaining two were for 911 calls for service to domestic disturbances. 

\section*{Discussion}
\label{sec:disc}

The results from this field study provide statistically significant evidence that a small-scale hot spot program based on proactive patrols, targeting only 0.5\% of Pittsburgh, leads to measurable reductions in serious violent crime in those areas, while avoiding over-policing arrests of racial minorities. These results contribute to the ongoing debate around proactive policing, showing both benefits and mitigation of certain potential harms when community-oriented policing practices are employed. However, determining whether to conduct hot spot policing in a given jurisdiction requires additional studies to assess the impact of proactive patrols on police-community relations and to rule out other potential harms to individuals related to more frequent interactions with police. If policy-makers assess that the benefits of such a program outweigh the risks, adequate oversight procedures should be created to ensure that proactive patrols follow best practices, avoiding aggressive and harmful policing practices, and to evaluate whether the program continues to achieve crime prevention benefits while avoiding unintended harms.

A reduction of 24 P1V crimes is 1.6\% of observed citywide violent crime during this period, corresponding to an estimated \$3.4 million in avoided crime costs to citizens and society. While the crime reduction is small with respect to citywide crime volume, it results from a relatively small amount of effort per patrol shift (an average of 35.5 minutes of hot spot patrol per day). It is reasonable to expect that hot spot-based policing would experience diminishing marginal benefits as the number of patrolled hot spots grows, but the small dosage of hot spot patrolling evaluated in this study and the steepness of the prediction tradeoff curve beyond 0.5 percent area on the horizontal axis nonetheless leave considerable capacity for scaling up the patrol program, as would eliminating car patrols and using foot patrols only. Additional studies are needed to evaluate the effects on crime volume of programs with different scales and to ensure that potential undesirable impacts related to over-policing are avoided as programs scale.

Chronic and temporary crime hot spots exhibit fundamental differences affecting the mechanisms by which proactive patrols deter crime. Chronic hot spots may experience elevated rates of crime for years or decades at a time. These highest-crime areas are typically known to police, but crime volume can remain high even in the presence of regular police patrols. By contrast, residential areas tend to experience low baseline rates of crime, but may represent areas where the fear of crime is highest \citep{Skogan1981, Moore1988}. Patrolling of temporary hot spots results in a spreading out of police resources to residential areas away from commercial zones where crime is chronically high. Results from this study indicate that patrols to temporary hot spots provide a meaningful reduction in crime counts, despite having significantly less overall crime volume than chronic hot spots.

This study provides evidence that hot-spot-based predictive policing has potential for substantial societal benefits through the prevention of serious violent crimes. However, predictive policing systems also present non-trivial risks of harms resulting from over-policing or under-policing, or potential exacerbation of societal inequalities. In \textit{SI: Best Practices for Hot Spot Prediction and Prevention}, we discuss a set of best practices for assessing and mitigating risks of bias and over-policing in predictive policing systems. These best practices include but are not limited to (1) using beneficial, community-oriented patrol protocols rather than aggressive enforcement, (2) appropriately choosing minimally-biased target variables and predictor variables for prediction models, (3) regularly vetting prediction models across multiple evaluation criteria, (4) ensuring geographic dispersion of targeted patrols with a dynamic prediction model, (5) considering impacts of the scale and intensity of a targeted intervention, and (6) designing a predictive policing system around place-based rather than person-based predictions. We encourage future researchers and users of predictive policing systems to carefully consider not only the potential impacts of a program on crime reduction, but also the other consequences, intended and unintended.

More empirical studies are needed to understand the potential impacts on crime volume and other citizen outcomes from hot spot programs of different scales and across cities. This study focuses on hot spot models that rely on reported crime data, and additional work is needed to understand the impacts of specific areas and communities being overrepresented or underrepresented in reported crime data. Further research is also needed to identify what patrol activities and strategies are most effective at fostering goodwill among the communities being policed in addition to providing crime reduction benefits. Ultimately, designing a predictive policing system that is both transparent and equitable is essential for long-term support from the general public.

\matmethods{\subsection*{Data}\label{sec:comparison_data} For the purpose of evaluating alternative crime forecast models and simulating a setting in which police command staff and crime analysts make weekly decisions about where to distribute proactive patrols, we relied only on data sources readily available to PBP for one-week ahead crime forecasts. Specifically, we obtained two data sets spanning five years of historical data from June 1, 2011 through June 1, 2016. The first data set is compiled from the City of Pittsburgh’s Automated Police Reporting System (APRS) and contains data on all 206,150 crime incidents recorded by the PBP within the five-year period of analysis. For each incident, the data set contains an associated crime code corresponding to the FBI Uniform Crime Reports (UCR) hierarchy. Criminal offenses are divided into two primary groups in the UCR hierarchy; Part I offenses represent the most serious categories of crime that are likely to be reported to police, and Part II crimes include less serious offenses. Part I crimes are further divided into two categories: violent crimes (P1V) and property crimes (P1P). SI Table S1 reports crime counts for the seven component crime types which make up all Part 1 offenses in the APRS data. (Though arson is also categorized by the FBI as a Part 1 offense, data on arson were not provided by PBP.)

Historical biases in policing activity disproportionately affect minority communities, and relying on historical crime reports could therefore result in a further entrenchment of these biases. Brantingham et al. \citep{Brantingham2017} examine whether biases in algorithmic place-based policing result in discriminatory consequences for minority groups and find no significant difference in arrest rates across racial-ethnic groups between treatment and control areas. Still, feedback loops are possible when relying on victimless crime data, such as drug arrests, for hot spot selection, in which case police have more discretion on enforcement than with crimes with victims \citep{Lum2016}. Areas may be initially selected for increased patrols due to high volume of historical victimless crime reports; these areas subsequently generate additional reports of crime due to increased police presence, then the same areas are again selected as hot spots, and so on. To mitigate this issue, we selected hot spots based on predictions of P1V crimes, which consist of serious violent offenses including homicide, rape, robbery, and aggravated assault. Reporting of these offenses is less likely than other crime types to depend on presence or absence of police in an area. Predictions of these crimes are therefore less prone to feedback loops resulting from increased patrols.

The second data set provided by PBP includes information on 911 calls for service to the Pittsburgh police, totaling approximately one million calls over the five-year period of analysis. This data includes the time and place that a call was made, as well as a code indicating the reason the call was made and a descriptive field indicating the outcome of the call. Taken together, the APRS data and CAD data provide a comprehensive picture of when and where crimes are being reported in Pittsburgh. An important caveat is that crimes that go unreported to police are not represented in these data. This analysis therefore evaluates prediction methods at forecasting crimes as they are reported to police, with the purpose of simulating one-week-ahead forecasts of reported crime made by police analysts.

\subsection*{Selection of Prediction Methods}
To understand the tradeoffs between various prediction methods and model specifications and to design an appropriate process for weekly hot spot selection, we evaluated the performance of a set of ten prediction models on the five years of historical crime data obtained from the PBP. Results from this analysis directly informed the choice and design of prediction models used in the field study of a hot-spot-based predictive policing program in Pittsburgh.

We divided historical APRS and CAD observations into a two-year model calibration period (June 1, 2011 through June 2, 2013) and a three-year evaluation period (June 3, 2013 through June 1, 2016). We conducted rolling one-week-ahead forecasts of P1V crime counts over the entire evaluation period to simulate crime analysts making weekly forecasts for hot spot selection. Input features provided to the models varied by model, but can include lagged counts of the target variable, and lagged counts of various ``leading indicator'' events which may be predictive of the outcome variable, namely other crime types or categories of 911 calls. Models were retrained throughout the evaluation period using a moving window of two years of training data, allowing models to adapt to changes in crime patterns to make optimal predictions. We compared ten prediction methods that have demonstrated success on forecasting crime counts or other types of spatiotemporal count data in previous studies: within-cell moving averages, kernel density estimation, L1-regularized logistic regression with lagged count features, L1-regularized logistic regression with crime cluster features, Gaussian process regression, univariate self-exciting point process modeling, self-exciting point process modeling with spatial covariates, multilayer perceptrons with lagged count features, multilayer perceptrons with lagged count and local neighborhood features, and convolutional neural networks. Models were evaluated on the prediction performance as well as equity in crime deterrence benefits, as measured by the spatial dispersion of top-predicted areas. A full discussion of the comparison of prediction methods is provided in \textit{SI: Evaluation of Prediction Methods on Historical Data}.

Based on the outcomes of the empirical comparison of prediction methods, we determined that using a single prediction model for hot spot selection would not perform well on the competing objectives of (1) prediction accuracy and (2) equitable distribution of hot spots throughout the city. A hot spot program with the sole aim of capturing the most crime within hot spots would likely be very static, with the same small set of chronic hot spots being selected for patrols each week. Based on the empirical results, we therefore chose separate prediction models for selecting weekly chronic and temporary hot spots for the field experiment. Chronic hot spot cells were selected using a cell-specific one-year moving average of crime counts. Temporary hot spot cells were selected using a fully connected feedforward neural network which predicts a difference from the long-term moving average within each cell. The temporary hot spot model includes time lags of target crimes, leading indicator crimes, and 911 call types (see SI Table S2 for a full list of leading indicator variables). Inputs to the hot spot models relied on the APRS and CAD data sets provided by PBP, refreshed on a weekly basis.

}

\showmatmethods{} % Display the Materials and Methods section

\acknow{We acknowledge the generosity of the Richard King Mellon Foundation for funding this study. We also acknowledge Chief Scott Schubert and Deputy Chief Thomas Stangrecki for their leadership in developing the hot spot program.
}

\showacknow{} % Display the acknowledgments section

\clearpage

\section*{Supplemental Information Appendix}

\section*{Evaluation of prediction methods on historical data}
\label{sec:comparison_results}

Table \ref{tab:crime_types} reports crime counts for the seven component crime types which make up all Part 1 offenses in data provided by the Pittsburgh Bureau of Police (PBP) from June 1, 2011 through June 1, 2016. Part 1 Violent (P1V) crimes were chosen as the target variable for prediction and model selection.

We compared ten prediction models that have demonstrated high performance on forecasting crime counts or other types of spatio-temporal count data in the literature. The training and prediction framework was standardized across the ten models, with model parameters tuned in a two-year calibration period from June 1, 2011 through June 2, 2013, and model performance for one-week-ahead predictions assessed in a three-year evaluation window from June 3, 2013 though May 29, 2016. Leading indicator variables for multivariate models are reported in Table \ref{tab:leading_indicators}. The models considered were: 

\begin{itemize}
  \item \textbf{Within-Cell Moving Average (MAVG)}. A window length of fixed size was selected to optimize prediction accuracy for the calibration period. At prediction time, target crimes are tabulated for each spatial unit over the entire time window ending at the present time period.
  \item \textbf{Kernel Density Estimation (KDE)}. A density surface of target crimes was estimated from spatial occurrences of the target variable over a fixed time window using a Gaussian kernel. Kernel bandwidth and time window size were chosen to optimize prediction accuracy for the calibration period.
  \item \textbf{Logistic regression with lagged count features and $L1$-regularization (LASSO-LC)}. A logistic regression model with $L1$-regularization was trained using lagged count features from leading indicator variables made up of individual crime types and 911 call categories. Counts were converted to binary class labels indicating presence or absence of crime, and the regularization parameter was selected to optimize prediction performance for the calibration period.
  \item \textbf{Logistic regression with crime cluster features and $L1$-regularization (LASSO-CC)}. Spatiotemporal clusters of leading indicator crimes were detected using the Fast Subset Scan approach of Neill \cite{Neill2012}. Cluster characteristics (size, duration, and intensity) were used as features for a sparse logistic regression model trained and tuned identically to LASSO-LC.
  \item \textbf{Gaussian Process Regression (GP)}. A Gaussian process regression model was trained assuming separable covariance across time and the two spatial dimensions. The isotropic squared exponential covariance function was applied for all three dimensions, and counts for leading indicators were included as linear terms in the mean function \citep{Rasmussen2005, Saatci2011}. GP hyperparameters were tuned to optimize prediction performance for the calibration period.
  \item \textbf{Univariate Self-Exciting Point Process (SEPP-UNI)}. As presented by Mohler \cite{Mohler2011}, a univariate self-exciting point process model was trained in which a set of background crimes of the target type are assumed to occur independently across time and space according to a stationary Poisson process, and subsequently result in elevated predicted risk for offspring events in the spatial vicinity. Background and offspring event rates are estimated iteratively through variable-bandwidth kernel density estimation.
  \item \textbf{Self-Exciting Point Process with spatial covariates (SEPP-MULTI)}. An extension of the SEPP-UNI model was trained to allow counts from leading indicator crimes to contribute to the overall intensity function of the target crime, following the training procedure described by Reinhart \cite{Reinhart2017}.
  \item \textbf{Multilayer Perceptron (MLP) with cell-specific lagged count features (MLP-LC)}. A densely-connected feedforward network was trained to predict target crime counts from 52 weeks of lagged crime data for a set of leading indicator crime types \citep{Olligschlaeger1998}. Model architecture was selected based on prediction accuracy in the calibration period, with a single hidden layer and ten hidden units outperforming other variants under consideration.
  \item \textbf{MLP with lagged count and local neighborhood features (MLP-NH)} Additional neighborhood features were added to the MLP-LC model, consisting of lagged crime counts for leading indicators within the eight cells (queen’s case) adjacent to the target cell for each observation.
 \item \textbf{Convolutional Neural Network (CNN)}. A convolutional neural network was adapted to predict target crime counts from a rectangular grid overlay of the city and lagged target crime counts within each cell. The CNN is designed to identify spatial patterns of target crimes which appear throughout the city and are predictive of future crime, rather than learning spatial structure individually within neighborhoods or other micro-areas.
\end{itemize}

A variety of evaluation metrics are available for measuring the relative and absolute performance of different prediction methods. For this analysis, the evaluation framework was designed to align closely with police goals for a hot-spot-based predictive policing program. In the place-based forecasting setting, a natural approach to assessing performance of crime prediction methods is to measure the prediction accuracy as the percent of crime volume predicted across a range of forecasted areas \citep{Gorr2015, Weisburd2009, Mohler2014}. A \textit{prediction tradeoff curve} plots the crime volume captured versus percentage area of a police jurisdiction forecasted over a range of area percentages from zero up to an upper limit determined by patrol resource constraints. The curve provides a visual depiction of the benefit/cost tradeoff faced by police when allocating resources to proactive patrols. If a single model outperforms others everywhere along the tradeoff curve (i.e., is the highest curve), then that model is likely to perform best regardless of how much area is ultimately selected for crime prevention.

We define a new measure, \textit{hot spot selection entropy}, to evaluate the spatial dispersion of top-predicted areas from each prediction model. For a given prediction model, let $n_i$ be the number of times grid cell $i$ appears in the top-predicted 1\% of all $N$ cells across all forecast periods, and define $p_i$ as the proportion of total hot spot selections occupied by cell $i$: 
\begin{equation*}
p_i = \frac{n_i}{\sum_{j=1}^N n_j}
\end{equation*}
Then hot spot selection entropy $H$ is calculated as the entropy of this hot spot distribution across grid cells:
\begin{equation*}
H = -\sum_{i=1}^N p_i \log_2 p_i
\end{equation*} 
\noindent Hot spot selection entropy is highest for predictions that lead to a uniform distribution of hot spots across the city, and lowest for prediction methods that choose the same hot spots in every time period. This statistic therefore provides a useful point of comparison relating to the geographic dispersion of hot spots selected by crime forecasting models.

\begin{figure}[h!]
  \setlength{\fboxsep}{0pt}
  \setlength{\fboxrule}{1pt}
 \centering

 \includegraphics[scale=0.3]{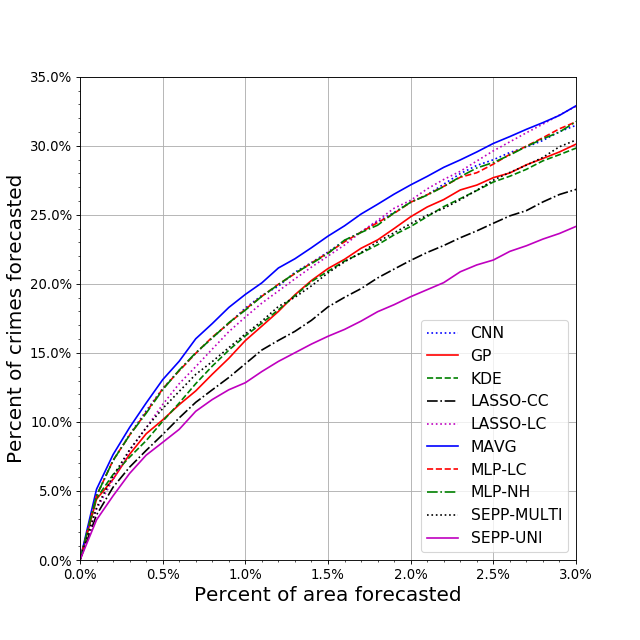}
\caption[Prediction tradeoff curves for 10 forecasting models on one-week-ahead predictions of Part 1 violent crime over evaluation period.]{Prediction tradeoff curves for 10 forecasting models on one-week-ahead predictions for Part 1 violent (P1V) crimes across a three year evaluation window.}
\label{fig:comparison_tradeoffs}
\end{figure}

Prediction tradeoff curves for all ten forecasting models on predicting P1V crimes are presented in Figure  \ref{fig:comparison_tradeoffs}. All models under consideration demonstrate some effectiveness at predicting crime, with all curves showing similar performance at the smallest levels of area forecasted within the city. As more area is forecasted, the relative performance of the models becomes clearer, with all models experiencing diminishing marginal rates of crime predicted as more area is forecasted. The MAVG model shows the best prediction performance in the top-predicted areas, from zero up to 3\% of the city area forecasted. Police departments are typically only able to effectively patrol 1-3\% of a city in a hot spot policing program, thus these results indicate the relatively simple MAVG method is likely the best candidate for selecting hot spots that capture the maximum amount of crime on average. With the exception of the relative poor performance of the SEPP-UNI and LASSO-CC models, most of the prediction models performed fairly similarly within the 0-3\% range of feasible areas shown in Figure \ref{fig:comparison_tradeoffs}. Although some of the models evaluated here are considerably more difficult to design, tune, and train than the MAVG model, it seems that the additional complexity of these models does not improve prediction performance within these feasible levels of patrol. 

Table \ref{tab:comparison_summary} reports summary statistics from the comparison of prediction methods using prediction curves of Figure \ref{fig:comparison_tradeoffs}. The partial area under the tradeoff curve (pAUC), analogous to the pAUC of receiver operating characteristic (ROC) curves, reports the area under the tradeoff curve up to 1\% of the city area forecasted. The pAUC results are a measure of predictive accuracy which relate directly to feasible levels of patrols that may be dispatched to the top 1\% of hot spots selected by these models. As depicted visually by the prediction tradeoff curves, predictions from the MAVG model give the highest pAUC at the 1\% level compared with all other models. This high level of prediction accuracy comes at the expense of decreased hot spot selection entropy (H) values. The MAVG model has among the lowest entropy values across models, indicating that the same cells are selected repeatedly in the top 1\% of predictions for this model. This lack of spatial dispersion of predictions is unsurprising given the design of the MAVG model, which leverages only long-term trends in crime to make predictions. The MLP-LC model provided the greatest variance in top-predicted cells, with hot spot selection entropy of 8.42 over the evaluation period. These results indicate that for a hot spot program where prediction accuracy and equity of proactive patrols are both key objectives, an approach that leverages multiple models is necessary to achieve good performance on both metrics simultaneously.

Table \ref{tab:percent_occupied} reports the percent of hot spot selections occupied by frequently-selected locations for a hot spot program targeting 1\% of a city’s total area, which provides an intuitive basis for understanding how frequently the top-selected hot spots change over time across models. Even the MLP-LC model, which had a high hot spot selection entropy relative to other methods, results in predicted hot spots that persist for long periods of time, with 43.6\% of hot spot selections occupied by locations that remain hot spots for greater than 75\% of the study period.

Through discussion with PBP command staff, we identified an objective to predict \textit{temporary hot spots}, which represent short-term flare-ups over baseline levels of crime. In order to predict temporary flare-ups in target crimes throughout the city, we modified the MLP-LC model by tweaking the target variable for prediction. MLP-DIFF is identical to MLP-LC in structure and inputs, but the outcome variable for prediction is changed from total observed crime counts to the difference between observed counts and a one-year moving average of target counts within each cell. The target variable is clipped at zero to prevent negative observed outcomes and predictions. The predictions from MLP-DIFF no longer represent the predicted number of target crime counts at each location, but instead represent the predicted positive difference from baseline levels of crime. Although the MLP-DIFF model results in lower prediction accuracy with respect to total crime captured, the hot spot selection entropy for this model is 10.53, much higher than the highest-entropy model considered in the previous section. Further, Table \ref{tab:percent_occupied} indicates that the MLP-DIFF model results in top-predicted areas persisting for much shorter periods of time than the MLP-LC and MAVG models.

Based on this evidence that the within-cell moving average (MAVG) and MLP-DIFF model address different police objectives for a proactive patrol program, we selected these two models for separate identification of chronic and temporary hot spots respectively. Despite the change in prediction interpretations for the MLP-DIFF model, we can still examine the prediction tradeoff curve and hot spot selection entropy from the temporary hot spot model to compare it with models predicting raw counts. In Figure \ref{fig:composite_tradeoff}, we show the tradeoff curves for the MAVG and MLP-DIFF models, as well as a composite curve that represents a combined chronic-temporary hot spots program with equal numbers of chronic (MAVG) and temporary (MLP-DIFF) hot spots. The composite curve is constructed to provide the tradeoff between the overall most accurate model, MAVG, and the composite model at any percentage area on the horizontal axis. From about 0.5\% to 1\% area, the vertical distance between the MAVG and composite curve is approximately constant, 1.5\% prediction accuracy, representing the prediction accuracy lost by adopting the composite curve instead of MAVG. Of course, the benefit from this loss is a substantial increase in spatial dispersion of crime prevention services by police.

\section*{Evaluation of hot spot program impacts} We report number of cell-weeks and counts and standard deviations of P1V crimes in treatment and control cells for chronic and temporary hot spots in Table \ref{tab:P1V_stdev}. Table \ref{tab:P1V_dose_reg} shows full regression results estimating dose-dependent treatment effects on P1V crimes in hot spots. We report results for four model variants: Model 1 examines the dose-dependent treatment effect of patrols to hot spot cells and includes week and zone-partition fixed effects. Model 2 additionally examines possible changes in crime volume in cells adjacent to hot spots and in all other cells (cells not selected as a hot spot and not adjacent to a hot spot). Model 3 estimates separate dose-dependent treatment effects for chronic and temporary hot spots. Finally, Model 4 estimates separate dose-dependent treatment effects for car patrols and foot patrols on chronic and temporary hot spots. The interaction terms with Dose variables represent the estimated dose-dependent treatment effect to hot spots, while the interaction terms with the Treatment variable represent the treatment effect of being in a treatment partition (either adjacent to or not adjacent to a hot spot).

We additionally provide non-dose-dependent regression results in Table \ref{tab:P1V_nondose_reg} estimating treatment effect of a cell being displayed to patrol officers as a hot spot, regardless of the observed number of proactive patrols to the location while selected as treatment. The interaction terms with the Treatment variable represent the treatment effect of being selected as a hot spot for hot spot cells, and the effect of being in a treatment partition for cells adjacent to or not adjacent to a hot spot. While P1V crimes represent the primary target for proactive patrols, we additionally provide P1P results in Table \ref{tab:P1P_nondose_reg} as evidence that crime prevention benefits are not limited to the targeted P1V crimes. Model 3 indicates that although proactive patrols targeted violent crime hot spots, patrols to chronic hot spots resulted in a statistically significant reduction in property crime in those areas. This effect on property crime was not observed in temporary hot spots, as these areas are less likely than chronic hot spots to be located in highly trafficked commercial areas where high levels of property crime tend to occur.

Table \ref{tab:cost_estimates} shows estimates of per-offense costs to society as reported by McCollister et al. \cite{McCollister2010}. Costs were inflated from 2008 dollars to 2018 dollars using the CPI inflation rate reported by the U.S. Bureau of Labor Statistics. We estimate the cost of proactive patrols overall to be less than \$300 thousand, as follows. The 2018 PBP budget for uniformed officers below the rank of sergeant was \$48.5 million for 755 officers with an average cost per officer of \$64,218, which we scale up by 20\% (to \$77,062) to additionally account for employment benefits provided to officers. Assuming 2,080 hours per officer per year, the cost per hour and officer was \$37.05. The hot spot program had 7,744 hours of patrol during the experimental year and thus cost \$286,906. The benefit/cost ratio was 11.9 (\$3,411,328 in benefits generated from \$286,906 in costs).

\section*{Equity in the allocation of crime prevention services by police}

The field of analytical ethics addresses what governmental programs should do in regard to equitably allocating scare resources. This section applies analytical ethics, as espoused by Anthony Culyer \cite{Culyer2001} in regard to governmental health care programs, to the case of crime prevention by municipal police. 

Culyer provides two preconditions for a governmental program to be considered for adoption. First, the program must be instrumental, a necessary condition if a more ultimate need is to be met. We argue that public safety, including crime prevention, is essential for the pursuit of happiness in other aspects of life as is health care. The field experiment of this paper provides statistical and practical evidence that the Pittsburgh hot spot program reduced serious violent crime. Second, only programs that are cost effective should be considered. The Pittsburgh program meets this requirement with a benefit/cost ratio of 11.9.

Culyer argues that need for a service is the most relevant factor for equitable allocation of scarce resources in a program. He provides two principles for meeting needs: “…that like needs should receive like attention and resources (horizontal equity) and greater needs should receive greater attention and resources (vertical equity).” Together these principles provide guidelines for the design of equitable programs given resource limitations. For example, in a crime hot spot program, some locations may need large allocations of patrol resources while other locations need only small allocations, and yet others may need none at all. A program that only targets some areas, or targets some areas substantially more often than others, could be considered “equitable” if its allocation of resources meets these needs. Nevertheless, two critical questions are left unanswered: how do we define need, and how should resources be allocated as a monotonically increasing function of need?

In the case of the hot spot program, a community’s need for proactive patrols refers primarily to reducing future instances of serious violent crime, yet should also consider the other potential impacts of policing on the well-being of the individuals being policed. While precisely quantifying need is challenging, it is clear that areas which would have a large number of serious violent crimes in the absence of crime prevention services are in greater need of such services. Thus we employ predictive models, estimating the expected number or probability of serious violent crimes that will occur in the next forecast period. To provide equitable crime prevention, we must first have models that are as accurate and unbiased as possible, necessitating our rigorous evaluation of multiple alternative models on historical data. We also quantified, through a field experiment, to what extent serious violent crimes would be reduced (and thus, needs met) through proactive patrols, showing that the proportional reduction is greater for temporary hot spots. Temporary hot spots also tend to target different (more residential and less commercial) and more dispersed areas than chronic hot spots, and thus might be preferred by policy makers for other reasons. This motivated us to provide two separate and largely independent forecast models for chronic and temporary hot spots as policy levers, allowing city policy makers to decide the number of chronic versus temporary hot spots when allocating crime prevention services. 

Given the prediction scores as an estimate of need, we must still consider how resources can be best allocated to achieve horizontal and vertical equity. In regard to horizontal equity, all areas of a city should be able to receive prevention resources whenever serious violent crimes are likely to occur in the near future, thus reducing the likelihood that such crimes will take place. In regard to vertical equity, more resources should be allocated to areas with higher prediction scores (i.e., higher estimated probability of future crimes), but it is clear that some such allocations are preferable to others. For example, at the small scale of the Pittsburgh program (0.5\% of the city with crime prevention services at a time), chronic hot spots are most effective in crime prevention. However, an all-chronic hot spot program concentrates police resources at the same few locations each week, leaving needs for the rest of the city unmet. Such an allocation is suboptimal with respect to horizontal equity, since there are many other areas which have nearly as high scores yet receive few or no crime prevention resources. Similarly, it is suboptimal with respect to vertical equity, in that only the few very highest-scoring areas are differentiated from other areas in their resource allocations. Our combined chronic and temporary hot spot program, while still focusing resources on areas with higher prediction scores, allocates resources to a much greater number of areas, as measured by the footprint map which shows all hot spots that received one week or more of crime prevention by police in a year. It also spreads resources more evenly between these areas, as shown by our entropy measure, thus improving both vertical and horizontal equity as compared to the all-chronic hot spot program.

Finally, we note that for large intervention doses such as saturation patrols, prediction scores might be insufficient to measure a community’s need for proactive policing: we must also consider the diminishing marginal returns of extremely frequent patrolling, as well as potential negative impacts due to over-policing. Given the much smaller intervention doses considered here, as well as the use of community policing to mitigate potential harms, we believe that allocating resources separately to chronic and temporary hot spots, as a monotonically increasing function of the prediction scores, is broadly consistent with Culyer’s definition of equity.

\section*{Best Practices for Crime Hot Spot Prediction and Prevention: Assessing and Mitigating Risks of Bias and Over-Policing}
The authors of this paper have listed below a number of recommended best practices based on our more than four decades of combined experience working with law enforcement and city leaders to develop and implement predictive hot-spot policing systems. These recommendations are supported by scientific evidence gathered from the field trial of predictive policing in Pittsburgh presented in the main text, as well as our expertise and understanding of the literature on policing and algorithmic fairness.

We believe that predictive policing has potential for substantial societal benefits through the prevention of serious violent crimes, but also presents non-trivial risks including the potential for (1) harms resulting from over-policing or under-policing of minority or low income areas, and (2) exacerbation of societal biases and inequalities, such as racial disparities in arrest rates. Considering these risks and benefits, we believe that the goals of predictive policing should be threefold: (1) reducing serious crimes, especially serious violent crimes as a first priority, that have the most substantial impacts on society; (2) improving equity in the distribution of policing resources; and (3) minimizing harms due to over-policing. Systems that only focus on the first goal run the risk of causing societal harms that outweigh their benefits, as well as negatively impacting already-disadvantaged neighborhoods and subpopulations.

We strongly encourage future researchers and users of predictive policing systems to adopt this broader view of what a predictive policing system should be designed to achieve, carefully considering not only the potential impacts of a program on crime reduction, but also the other consequences, both intended and unintended, of such systems. Such consideration should have two aspects: (1) incorporation of best practices for system design, implementation, and deployment, and (2) continued measurement of societal impacts as a system operates, with iterative improvement of both the system and corresponding policing practices to maximize benefits, minimize harms, and improve equity.

\subsection{Use beneficial rather than aggressive policing interventions}
Perhaps the most critical factor influencing the societal impacts of a predictive policing program is the policing intervention itself: when an area is targeted for a hot-spot patrol, how do the patrolling officers interact with citizens in that area? Aggressive policing practices such as broken-windows policing, which includes strict enforcement of low-level crimes (leading to increased arrests for minor offenses such as marijuana possession and loitering) and frequent use of pedestrian stops, has been shown to cause substantial harms to disadvantaged and minority communities, including adverse impacts on health and educational outcomes \citep{Legewie2019}. In our field trial of predictive policing in Pittsburgh, we reaffirmed the Pittsburgh Bureau of Police’s policy that officers use community-oriented policing while on proactive patrol. In community-oriented policing, officers focus on problem solving and citizen engagement (via foot patrols), for example, warning of possible violent crimes and gathering information. The result was that, during our study, very few arrests were made by hot spot officers while on patrol, nor were arrests increased for commonly over-policed crimes.

We strongly encourage future users of predictive policing systems to think carefully about the way in which they intervene, to avoid aggressive policing approaches that cause detrimental outcomes to already vulnerable communities, and to ensure that all hot-spot patrol officers are properly trained in community-oriented policing techniques.

\subsection{Appropriately choose the target variable for prediction}
In our field trial of predictive policing in Pittsburgh, we focused on predicting and preventing Part 1 Violent crimes: the serious crimes of homicide, rape, robbery, and aggravated assault. In addition to representing the types of crime which were of greatest interest to Pittsburgh law enforcement officials and city leaders, and those with the greatest costs to society (as estimated by McCollister et al. \cite{McCollister2010}), the choice of serious violent crimes as the target variable also mitigates potential biases in several ways. With the exception of domestic crimes (not considered in our study), serious violent crimes tend to be reported to the police \citep{BJS2019}, as opposed to victimless crimes such as drug offenses, which tend to be discovered only if officers are looking for them.

The results of this are twofold: first, arrests for minor crimes, and particularly victimless crimes, tend to exhibit strong racial biases, for example, from 2001-2010, African-Americans were arrested 3.73 times as often as whites for marijuana possession despite approximately equal rates of use \citep{ACLU2013}. Thus targeting of such crimes for prevention, in addition to being an inefficient use of police resources, exacerbates existing societal biases including disparities induced by current policing practices. Second, as argued by Lum and Isaac \cite{Lum2016}, targeting victimless crimes has high potential to lead to undesirable feedback loops, where policing in an area discovers the presence of victimless crimes, which are then used to justify sending additional patrols to that area, resulting in rapid amplification of disparities. Thus, we strongly recommend targeting serious violent crimes, or a similar quantity, rather than minor and victimless crimes.

\subsection{Appropriately choose predictor variables to minimize bias}
A number of variables commonly used as predictors in hot-spot policing have the potential to induce biases into the predictive model, and should be avoided. As noted above, crime offense data for victimless crimes such as drug possession have the potential to be racially biased because of racial disparities in arrest rates. We advise against using arrest data, which is likewise biased by policing practices. Furthermore, using demographic data (such as race), socio-economic data (such as poverty level), or neighborhood-level proxies for race and socio-economic status as predictors should be avoided; we recommend using demographic data only to evaluate systems and to determine whether they have had disparate impacts on poor and minority communities. In our randomized field trial of predictive policing in Pittsburgh, our hot spot prediction models do not include socio-economic, racial, individual person, drug possession or dealing offense, or arrest data; they only use selected crime offense and 911 emergency call data. Crime offenses include all incidents in which a crime is reported, whether or not there was an associated arrest. The chosen sources of data are likely to be less biased than those for arrests and victimless crimes, but are not necessarily unbiased; thus, whenever possible, additional sources of bias (such as reporting biases in 911 call data) should be assessed and mitigated.

\subsection{Regularly vet prediction models on multiple evaluation criteria} Selecting a prediction model for identification of hot spots requires evaluating candidate models on each of the multiple objectives of a predictive policing program. First and foremost, predictions should reflect actual underlying crime patterns, with selected hot spots capturing meaningful proportions of total crime. Prediction tradeoff curves show the full tradeoff between crime volume captured by predicted hot spots and area treated by a hot-spot policing program, enabling comparison of models on predicted crime volume across hot-spot programs of varying scales. Candidate models may vary considerably in their mechanism of prediction, with some capturing long-term crime trends and others responding to short-term shifts in crime patterns. Measuring the geographic dispersion of predicted hot spots across forecast periods indicates how dynamic a hot-spot policing program will be, and whether a given prediction model can equitably distribute police effort across areas that could benefit from additional community-oriented policing. Finally, the demographic and neighborhood characteristics of hot spots from candidate prediction models indicate which models might contribute to over-policing and provide critical context for designing beneficial policing interventions.

Crime patterns shift over time and in response to police interventions. Any operational prediction model(s) underlying a hot-spot policing program should be periodically reevaluated on the full set of objectives outlined above, to ensure that predicted hot spots are continuing to provide intended benefits. Ideally, regular vetting of operational models is conducted by both (1) analysts with technical proficiency in predictive modeling, and (2) law enforcement officers with awareness of local crime and neighborhood characteristics. Leveraging expertise in both the technical underpinnings of hot-spot prediction and local crime patterns enables periodic course correction to improve and maintain effectiveness of operational models over time.

\subsection{Ensure the geographic dispersion of targeted patrols; use chronic and temporary hot spots to balance efficacy and equity} We believe that geographic dispersion of patrols, spreading police presence more equitably according to need throughout the city, is a necessary component of equity in policing. Placing too many police resources in a given area has the potential to harm both that area (due to risks of over-policing) as well as other areas (which are not receiving sufficient resources for prevention of serious violent crimes).

For our field study of predictive policing in Pittsburgh, conscious of the need to balance crime prevention with equitable resource distribution while avoiding over-policing, we assessed the impacts of patrolling two different types of crime hot spots, chronic and temporary. As compared to chronic hot spots, temporary hot spot policing improved simple measures of equity and avoidance of over-policing: we observed much greater dispersion of patrols across the city, and each temporary hot spot was treated no more than 15\% of the time over the 52-week study period. Dispersion of patrols was substantially lower for chronic hot spots, and some areas would be targeted essentially every week by an operational chronic hot spot program without randomization. (In our cross-over experimental design, chronic hot spots had the maximum potential of being targeted for patrols half the time.)

The desired trade-off between temporary and chronic hot spots is not entirely clear: temporary hot spots increase equity, but their effectiveness may not scale to larger predictive policing programs. More generally, both temporary and chronic hot spots seem to be potentially useful targets for predictive policing, for small-scale programs with attention paid to equity. In addition to the use of temporary hot spots to spread prevention benefits across the city, we encourage randomization of patrols among chronic hot spots with similar predicted probabilities, to reduce disparities between communities with similar policing needs. 

\subsection{Consider impacts of the intensity and scale of the targeted intervention} 
\begin{figure}[h!]
  \setlength{\fboxsep}{0pt}
  \setlength{\fboxrule}{1pt}
 \centering
 \includegraphics[scale=0.3]{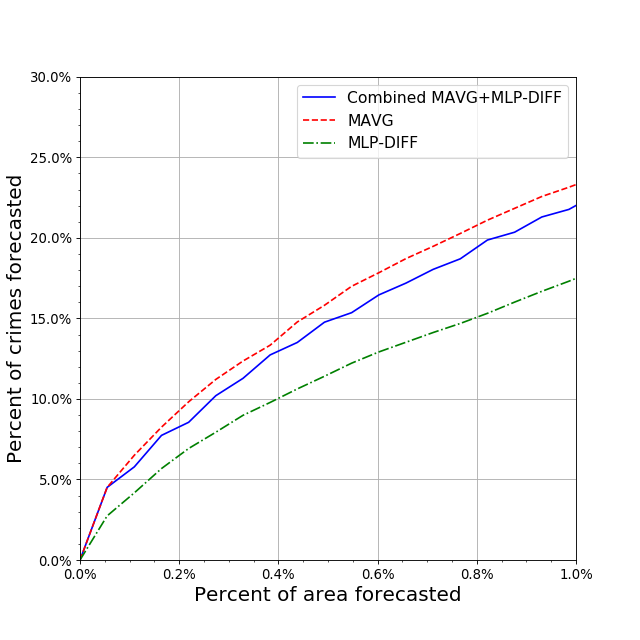}
\caption[Composite tradeoff curve representing a combined chronic-temporary hot spot program.]{Composite tradeoff curve representing a combined chronic-temporary hot spot program.}
\label{fig:composite_tradeoff}
\end{figure}

Our randomized field trial of predictive policing in Pittsburgh demonstrates the potential to achieve small but significant crime reductions through a small-scale, targeted policing program. So one might consider trying to gain more crime reduction through a larger-scale program (intervening on more hot spots) and/or through higher patrol intensities (targeting more patrols per hot spot). However, there are several reasons to carefully assess societal costs and benefits before scaling up. First, a larger-scale program may simply be infeasible given police resources. In our Pittsburgh study, which targeted patrols using roughly half of patrol officers’ discretionary time, our desired intensity of treatment was to average one additional 15-minute community-oriented patrol per 8-hour police shift, but patrol intensity averaged 79\% of target patrol duration (and 62\% of target patrol counts) over the study period. In this setting, increasing the number of hot spots would spread resources too thinly to achieve the desired prevention effect. Second, even given sufficient police resources, increasing the number of hot spots will likely require targeting areas with lower predicted crime rates than the “hottest” hot spots, resulting in fewer crimes prevented per hot spot. We note that the tradeoff curves of Figures \ref{fig:comparison_tradeoffs} and \ref{fig:composite_tradeoff} are still fairly steep beyond 0.5\% area, both for the MAVG model for chronic hot spots and for the combined model, indicating some capacity to scale up the program. Furthermore, given that the field study showed that only foot patrols (not car patrols) significantly reduce P1V, the possibility exists of increasing effectiveness with the same level of patrols by using only foot patrols. Third, increasing the number of hot spots may decrease the accuracy of crime prediction, particularly for temporary hot spots. Fourth, increasing the number of hot spots without explicitly considering geographic dispersion increases the likelihood that any given area will be repeatedly targeted with patrols, thus increasing the risk of negative impacts due to over-policing. Similarly, increasing patrol intensities in a given area would be expected to have diminishing marginal returns (the impact of adding an extra unit decreases as more crime is prevented) as well as increasing marginal costs (both due to risks of over-policing as well as increased disparities between targeted and non-targeted areas).

\subsection{Avoid introducing inequalities via experimental design}
When conducting controlled experiments to assess the impacts of new policing practices, including predictive policing, inequality could arise from the long-term, arbitrary assignment of some areas to a treatment group and some areas to a control group, including but not limited to over-policing of the treatment group and/or under-policing of the control group. Thus we recommend the use of cross-over experimental designs whenever possible to balance the total amount of targeted policing effort across the entire study area. In this study design, each zone is divided into subareas assigned to, say, group A and B, of roughly equal crime levels for the crimes being prevented. The subareas can number more than two and should be bounded by locations with low crime rates to protect against spatial contamination of control hot spots by nearby treatment hot spots. Then in subsequent periods, or selected randomly, A and B alternate being treatment versus control areas for locations of hot spots. Generally, the effect of treatment on a hot spot has a residual period on the order of hours after patrol, so that temporal contamination is not an issue between time periods. If, however, there is a general treatment effect lasting more than a period of treatment, the cross-over design fails. We believe that a general effect will not take place during an experimental period, but could be possible for extended periods of treatment.

\subsection{Systematically measure impacts, both intended and unintended}
For operational deployment of predictive policing systems, continued measurement of impacts on the policed communities is essential to assess and mitigate any unintended consequences of targeted patrolling. Ideally, this assessment should be conducted by an organization that is impartial and independent of the law enforcement agency conducting predictive policing, with results supplied both to the Mayor’s Office and the public (at regular intervals, to improve transparency) and to law enforcement (for continuous performance improvement).

Quantities to measure could include arrests (particularly those for commonly over-policed crimes and minor drug offenses), police use of force, citizen contacts, and crime victimization, and should be broken down both by neighborhood and by demographics (e.g., to assess racial disparities in policing). Both arrests made by hot-spot officers while on targeted patrols, as well as overall arrests, should be measured, to assess both direct and indirect impacts of the program. In addition to measuring the number of observed crimes in the targeted areas, crime displacement to nearby areas should also be assessed, though we did not observe any displacement of crime in our Pittsburgh study. Finally, we encourage continued measurement of the number of patrols each hot spot receives each week, both to keep track of treatment intensity (and assess its potential impacts on both crime reduction and arrests) and to ensure that an adequate number of patrols are conducted to achieve crime prevention benefits.

\subsection{Predict where, not who}
While the focus of this discussion is on best practices for geographically targeted prediction and prevention of crime, some cities have instead focused on crime prediction at the individual level. For example, the City of Chicago’s Strategic Subjects List predicts individuals’ probability of being involved in a shooting, either as perpetrator or victim. We believe that, in addition to sharing the numerous challenges of place-based crime forecasting, person-based forecasting creates substantial privacy concerns as well as presenting increased risk of harm to individuals and communities, for example, due to false positives. As we have argued in \citep{Fitzpatrick2019}, ``Person-based forecasting identifies specific individuals as posing a high risk for crime, some of whom may have never committed a crime in the past. Place-based predictions offer less opportunity for officers to make biased judgments about an individual’s propensity for crime or violence prior to engagement. Such biases can lead to poor decision-making and improper use of force.''

\begin{table}\centering
\caption{Counts of Automated Police Reporting System crime incidents for all Part 1 offenses from June 1, 2011 through June 1, 2016 in Pittsburgh, PA. Uniform Crime Reports (UCR) codes correspond to a hierarchy of crime types employed by the Federal Bureau of Investigation.}
\begin{tabular}{l|c|l|c}
\toprule
&UCR Code &Crime Type & Frequency\\
\cline{2-4}
\multirow{4}{*}{\shortstack{Part 1 Violent\\ (P1V) Offenses}} 
&  1 & Criminal Homicide & 296\\\cline{2-4}
&  2 & Forcible Rape & 991\\\cline{2-4}
&  3 & Robbery & 5822\\\cline{2-4}
&  4 & Aggravated Assault & 5856\\
\midrule
\multirow{3}{*}{\shortstack{Part 1 Property\\ (P1P) Offenses}} 
&  5 & Burglary & 11665\\\cline{2-4}
&  6 & Larceny/Theft & 37230\\\cline{2-4}
&  7 & Vehicle Theft & 3870\\
\bottomrule
\end{tabular}
\label{tab:crime_types}
\end{table}

\begin{table}\centering
\caption{Leading indicator variables provided as inputs to multivariate prediction models.}
\begin{tabular}{l|l}
\toprule
Data Source&Leading Indicator Variable\\
\cline{1-2}
\multirow{14}{*}{\shortstack{Automated Police\\Reporting System\\ (APRS)}} 
& Criminal Homicide\\\cline{2-2}
& Forcible Rape\\\cline{2-2}
& Robbery\\\cline{2-2}
& Aggravated Assault\\\cline{2-2}
& Burglary\\\cline{2-2}
& Larceny/Theft\\\cline{2-2}
& Vehicle Theft\\\cline{2-2}
& Simple Assault\\\cline{2-2}
& Vandalism\\\cline{2-2}
& Liquor Law Violations\\\cline{2-2}
& Public Drunkenness\\\cline{2-2}
& Disorderly Conduct\\\cline{2-2}
& Criminal Mischief\\\cline{2-2}
& Trespass\\
\midrule
\multirow{11}{*}{\shortstack{Computer-Aided\\Dispatch (CAD)}} 
& Assault\\\cline{2-2}
& Burglary\\\cline{2-2}
& Criminal Mischief\\\cline{2-2}
& Disorderly Person\\\cline{2-2}
& Disturbance\\\cline{2-2}
& Drug-related Complaint\\\cline{2-2}
& Harassment\\\cline{2-2}
& Larceny\\\cline{2-2}
& Suspicious Activity\\\cline{2-2}
& Vehicle Theft\\\cline{2-2}
& Weapons or Gunshot Complaint\\
\bottomrule
\end{tabular}
\label{tab:leading_indicators}
\end{table}

\begin{table}\centering
\caption{Partial area under the tradeoff curve ($pAUC$) and hot spot selection entropy ($H$) for a 1\% target area from 10 forecasting models across a three year evaluation window. Best-performing model for each column denoted in bold face.}
\begin{threeparttable}
\begin{tabular}{lc|c|}
\toprule
& \multicolumn{2}{c}{Part 1 violent crimes}\\
\cmidrule(rl){2-3}
&$pAUC \times10^2$& $H$\\
\midrule
MAVG&\textbf{1.31}&6.38  \\\hline
MLP-LC &1.23&\textbf{8.42}\\\hline
MLP-NH&1.23&8.21 \\\hline
CNN &1.23& 7.43\\\hline
LASSO-LC&1.14&7.06  \\\hline
SEPP-MULTI&1.10&7.58  \\\hline
KDE &1.06&6.83\\\hline
GP&1.05&6.30 \\\hline
LASSO-CC&0.94&8.31\\\hline
SEPP-UNI &0.87&6.34\\
\bottomrule
\end{tabular}
% \begin{tablenotes}
%       \small
%       \item \textbf{Best-performing model for each column denoted in bold face.}
%     \end{tablenotes}
  \end{threeparttable}
\label{tab:comparison_summary}
\end{table}

\begin{table*}\centering
\caption{Percent of hot spot selections occupied by frequently-selected locations (1\% target area).}
\begin{tabular}{lc|c|c|c}
\toprule
& \multicolumn{4}{c}{\% of Total Hot Spot Selections Occupied by Locations Persisting For:}\\
\cmidrule(rl){2-5}
&$>$ 25\% of All Weeks& $>$ 50\% of All Weeks&$>$ 75\% of All Weeks&100\% of All Weeks\\
\midrule
MAVG&96.2\%&91.0\%  &82.6\%&69.6\%  \\\hline
MLP-LC &74.2\%&55.5\%  &43.6\%&0.0\%  \\\hline
MLP-DIFF &3.5\%&1.7\%  &0.0\%&0.0\%  \\
\bottomrule
\end{tabular}
\label{tab:percent_occupied}
\end{table*}

\begin{table*}\centering
\caption{Counts and standard deviations of P1V crimes in treatment and control hot spots across cell-weeks.}
\begin{tabular}{l|c|c|c|c|c|c|c}
\toprule
& \multicolumn{3}{c|}{Control}&\multicolumn{3}{c|}{Treatment}&\% Reduction\\\cmidrule(rl){2-7}
&Sum of P1V& Std. Dev.&N (cell-weeks)&Sum of P1V& Std. Dev.&N (cell-weeks)&in P1V\\
\midrule
All Hot Spots&95&0.22&1872&71&0.20&1872&25.3\%\\\hline
\hspace{5pt}Chronic Hot Spots&80&0.29&936&61&0.26&936&23.8\%\\\hline
\hspace{5pt}Temp. Hot Spots&15&0.13&936&10&0.10&936&33.3\%\\\hline
Adjacent Cells&163&0.12&11869&149&0.11&12135&8.6\%\\
\bottomrule
\end{tabular}
\label{tab:P1V_stdev}
\end{table*}

\begin{table*}
\centering
\begin{threeparttable}
\caption[Regression results estimating dose-dependent treatment effect on Part 1 violent crime counts.]{Ordinary least squares (OLS) regression results estimating dose-dependent treatment effect of proactive patrols on P1V crime counts. Units of analysis are cell-weeks. Reference group for Model 1 is non-hot spot cells. Reference group for Models 2-4 is control cells not selected as hot spots or adjacent to hot spots.}
\label{tab:P1V_dose_reg}
\def\arraystretch{0.9}
\begin{tabular}{lcccc}
\toprule
 & (1)& (2)&(3)&(4)\\
\midrule
Hot Spot       & 0.0431***  &0.0437***            & &\\
                     & (0.040, 0.046)    & (0.041,0.046) & &\\
Hot Spot $\times$ Dose (Combined)  &-0.0009***& -0.0009*** & &\\
              & (-0.001, 0.000)    & (-0.001, 0.000) & &\\
Chronic Hot Spot     & & &0.0772*** &0.0768***\\& & & (0.074, 0.081)&(0.073, 0.080)\\
Chronic Hot Spot $\times$ Dose (Combined)    &  & &   -0.0015*** \\& & & (-0.002, -0.001)&\\
Chronic Hot Spot $\times$ Dose (Car)   &  & &   &0.0002 \\& & & &(0.000, 0.001)\\
Chronic Hot Spot $\times$ Dose (Foot)    &  & &   &-0.0031*** \\& & & &(-0.005, -0.002)\\
Temp. Hot Spot     & & & 0.0122*** &0.0116***\\ & && (0.009, 0.016)&(0.008, 0.015)\\
Temp. Hot Spot $\times$ Dose (Combined)     &  & &  -0.0008**&\\& & & (-0.001, 0.000)&\\
Temp. Hot Spot $\times$ Dose (Car)    &  & &  &0.0006*  \\& & & &(0.000, 0.001)\\
Temp. Hot Spot $\times$ Dose (Foot)     &  & &  &-0.0025***  \\& & & &(-0.004, -0.001)\\
Adj. to Hot Spot    && 0.0102*** & 0.0102*** &0.0102***\\
              & & (0.009, 0.011)& (0.009, 0.011)&(0.009, 0.011)\\
(Adj. to Hot Spot) $\times$ Treatment     && -0.0014* & -0.0014* &-0.0014* \\
              & & (-0.003, 0.000)& (-0.003, 0.000)&(-0.003, 0.000)\\
(Not Adj. to Hot Spot) $\times$ Treatment    && -0.0001 & -0.0001 & -0.0001 \\
              & &(-0.001, 0.000) & (-0.001, 0.000) &(-0.001, 0.000)\\
Constant      & 0.0041*** & 0.0035***            &0.0035*** & 0.0035*** \\
              & (0.002, 0.006)       & (0.002, 0.005) & (0.002, 0.005)&(0.002, 0.005)\\
Week FE & Included&Included&Included&Included\\
Zone-Partition FE &Included&Included&Included&Included\\
\midrule
Adj. $R^2$&0.005&0.007&0.009&0.009\\
No. Observations&361719&361719&361719&361719\\
\bottomrule
\end{tabular}
\begin{tablenotes}
      \small
      \item Notes: 95\% confidence intervals shown in parentheses. Significance at the 1\% level is denoted by ***; ** denotes significance at the 5\% level; and * significance at the 10\% level.
    \end{tablenotes}
  \end{threeparttable}
\end{table*}

\begin{table*}
\centering
\begin{threeparttable}
\caption[Regression results estimating non-dose-dependent treatment effect on Part 1 violent crime counts.]{Ordinary least squares (OLS) regression estimating non-dose-dependent treatment effect on Part 1 Violent crime counts. Units of analysis are cell-weeks. Reference group for Model 1 is non-hot spot cells. Reference group for Models 2 and 3 is control cells not selected as hot spots or adjacent to hot spots.}
\label{tab:P1V_nondose_reg}
\def\arraystretch{0.9}
\begin{tabular}{lccc}
\toprule
 & (1)& (2)&(3)\\
\midrule
Hot Spot & 0.0465***  &0.0472***            &  \\
                     & (0.044, 0.049)    & (0.044, 0.050) & \\
Hot Spot $\times$ Treatment    &-0.0128***& -0.0128*** & \\
              & (-0.017, -0.009)    & (-0.017, -0.009) & \\
Chronic Hot Spot     & & &0.0818*** \\& & & (0.078, 0.086)\\
Chronic Hot Spot $\times$ Treatment     &  & &   -0.0203*** \\& & & (-0.026, -0.014)\\
Temp. Hot Spot     & & & 0.0125*** \\ & && (0.008, 0.017)\\
Temp. Hot Spot $\times$ Treatment      &  & &  -0.0053*  \\& & & (-0.011, 0.000)\\
Adj. to Hot Spot      && 0.0102*** & 0.0102*** \\
              & & (0.009, 0.011)& (0.009, 0.011)\\
(Adj. to Hot Spot) $\times$ Treatment     && -0.0014* & -0.0014* \\
              & & (-0.003, 0.000)& (-0.003, 0.000)\\
(Not Adj. to Hot Spot) $\times$ Treatment     && -0.0001 & -0.0001 \\
              & &(-0.001, 0.000) & (-0.001, 0.000) \\
Constant      & 0.0041*** & 0.0035***            &0.0035***  \\
              	    & (0.002, 0.006)       & (0.002, 0.005) & (0.002, 0.005)\\
Week FE & Included&Included&Included\\
Zone-Partition FE &Included&Included&Included\\
\midrule
Adj. $R^2$&0.005&0.007&0.009\\
No. Observations&361719&361719&361719\\
\bottomrule
\end{tabular}
\begin{tablenotes}
      \small
      \item Notes: 95\% confidence intervals shown in parentheses. Significance at the 1\% level is denoted by ***; ** denotes significance at the 5\% level; and * significance at the 10\% level.
    \end{tablenotes}
  \end{threeparttable}
\end{table*}

\begin{table*}
\centering
\begin{threeparttable}
\caption[Regression results estimating non-dose-dependent treatment effect on Part 1 property crime counts.]{Ordinary least squares (OLS) regression estimating non-dose-dependent treatment effect on Part 1 Property crime counts. Units of analysis are cell-weeks. Reference group for Model 1 is non-hot spot cells. Reference group for Models 2 and 3 is control cells not selected as hot spots or adjacent to hot spots.}
\label{tab:P1P_nondose_reg}
\def\arraystretch{0.9}
\begin{tabular}{lccc}
\toprule
 & (1)& (2)&(3)\\
\midrule
Hot Spot & 0.1501***  &0.1525***            &  \\
                     & (0.142, 0.158)    & (0.145, 0.160) & \\
Hot Spot $\times$ Treatment    &-0.0093*& -0.0093* & \\
              & (-0.020, 0.001)    & (-0.020, 0.001) & \\
Chronic Hot Spot     & & &0.2356*** \\& & & (0.225, 0.246)\\
Chronic Hot Spot $\times$ Treatment     &  & &   -0.0157** \\& & & (-0.031, -0.000)\\
Temp. Hot Spot     & & & 0.0693*** \\ & && (0.058, 0.080)\\
Temp. Hot Spot $\times$ Treatment      &  & &  -0.0029  \\& & & (-0.018, 0.012)\\
Adj. to Hot Spot      && 0.0342*** & 0.0342*** \\
              & & (0.031, 0.037)& (0.031, 0.037)\\
(Adj. to Hot Spot) $\times$ Treatment     && -0.0034 & -0.0034 \\
              & & (-0.008, 0.001)& (-0.008, 0.001)\\
(Not Adj. to Hot Spot) $\times$ Treatment     && 0.0001 & 0.0001 \\
              & &(-0.001, 0.001) & (-0.001, 0.001) \\
Constant      & 0.0171*** & 0.0150***            &0.0150***  \\
              	    & (0.013, 0.021)       & (0.011, 0.019) & (0.011, 0.019)\\
Week FE & Included&Included&Included\\
Zone-Partition FE &Included&Included&Included\\
\midrule
Adj. $R^2$&0.012&0.014&0.016\\
No. Observations&361719&361719&361719\\
\bottomrule
\end{tabular}
\begin{tablenotes}
      \small
      \item Notes: 95\% confidence intervals shown in parentheses. Significance at the 1\% level is denoted by ***; ** denotes significance at the 5\% level; and * significance at the 10\% level.
    \end{tablenotes}
  \end{threeparttable}
\end{table*}

\begin{table}[h!]\centering
  \begin{threeparttable}
\caption[Per-offense estimates of costs to society from Part 1 crime types.]{Per-offense estimates of costs to society from Part 1 crime types.}
\label{tab:cost_estimates}
\begin{tabular}{c|l|c|c}
\toprule
\multicolumn{2}{c|}{}&Cost* (\$2008) & Cost** (\$2018)\\
\midrule
\multirow{4}{*}{Part 1 Violent}
&Murder/Manslaughter&\$8,982,907&\$10,548,448\\\cline{2-4}
&Forcible Rape &\$240,776&\$282,738\\\cline{2-4}
&Aggravated Assault&\$107,020 &\$125,671\\\cline{2-4}
&Robbery&\$42,310 &\$49,684\\
\midrule
\multirow{3}{*}{Part 1 Property}
&Burglary&\$6,462&\$7,588\\\cline{2-4}
&Larceny &\$3,532 &\$4,148\\\cline{2-4}
&Vehicle Theft&\$10,772&\$12,649   \\
\bottomrule
\end{tabular}
\begin{tablenotes}
      \small
      \item *Reported in McCollister et al. \cite{McCollister2010}.
      \item  **Costs in 2008 dollars were inflated to 2018 dollars using \\\url{https://www.bls.gov/data/inflation_calculator.htm}.
    \end{tablenotes}
  \end{threeparttable}
\end{table}

\section*{References}
\bibliographystyle{pnas-new}
\bibliography{hotspot_paper}

\end{document}